\documentclass[pdflatex,sn-mathphys]{sn-jnl}

\usepackage{subcaption}

\usepackage{float}

\jyear{2022}%

\begin{document}
\title[ML Extraction of BCs from Doppler Echo Images for Patient Specific CoA:  CFD Study]{Machine Learning based Extraction of Boundary Conditions from Doppler Echo Images for Patient Specific Coarctation of the Aorta:  Computational Fluid Dynamics Study}


\author*[1]{\fnm{Vincent Milimo Masilokwa} \sur{Punabantu}}\email{vincentvmp@gmail.com}

\author[1]{\fnm{Malebogo} \sur{Ngoepe}}\email{malebogo.ngoepe@myuct.co.za}

\author[2]{\fnm{Amit Kumar} \sur{Mishra}}\email{akmishra@ieee.org}

\author[3]{\fnm{Thomas} \sur{Aldersley}}\email{thomas.aldersley@uct.ac.za}

\author[3]{\fnm{John} \sur{Lawrenson}}\email{john.lawrenson@uct.ac.za}

\author[4]{\fnm{Liesl} \sur{Zuhlke}}\email{liesl.zuhlke@uct.ac.za}

\affil*[1]{\orgdiv{Mechanical Engineering}, \orgname{University of Cape Town}, \city{Cape Town}, \postcode{7700}, \state{Western Cape}, \country{South Africa}}

\affil[2]{\orgdiv{Electrical Engineering}, \orgname{University of Cape Town}, \city{Cape Town}, \postcode{7700}, \state{Western Cape}, \country{South Africa}}

\affil[3]{\orgdiv{The Children's Heart Disease Research Unit}, \orgname{Red Cross Children's Hospital}, \city{Cape Town}, \postcode{7700}, \state{Western Cape}, \country{South Africa}}

\affil[4]{\orgdiv{Extramural Research and Internal Portfolio}, \orgname{South African Medical Research Council}, \country{South Africa}}


\abstract{
Purpose- Coarctation of the Aorta (CoA) patient-specific computational fluid dynamics (CFD) studies in resource constrained settings are limited by the available imaging modalities for geometry and velocity data acquisition. Doppler echocardiography has been seen as a suitable velocity acquisition modality due to its higher availability and safety. This study aimed to investigate the application of classical machine learning (ML) methods to create an adequate and robust approach for obtaining boundary conditions (BCs) from Doppler Echocardiography images, for haemodynamic modelling using CFD. 

Methods- Our proposed approach combines ML and CFD to model haemodynamic flow within the region of interest. With the key feature of the approach being the use of ML models to calibrate the inlet and outlet boundary conditions (BCs) of the CFD model. The key input variable for the ML model was the patients heart rate as this was the parameter that varied in time across the measured vessels within the study. ANSYS Fluent was used for the CFD component of the study whilst the scikit-learn python library was used for the ML component.

Results- We validated our approach against a real clinical case of severe CoA before intervention. The maximum coarctation velocity of our simulations were compared to the measured maximum coarctation velocity obtained from the patient whose geometry is used within the study. Of the 5 ML models used to obtain BCs the top model was within  5\% of the measured maximum coarctation velocity.

Conclusion- The framework demonstrated that it was capable of taking variations of the patients heart rate between measurements into account. Thus, enabling the calculation of BCs that were physiologically realistic when the heart rate was scaled across each vessel whilst providing a  reasonably accurate solution.
}




\maketitle
\section{Introduction}\label{sec1}



Congenital  heart  disease  (CHD)  is  the  most  common  birth  defect,  with  a global prevalence  of approximately  9  per  1000  births. Coarctation  of the  aorta  (CoA) is one  of  the  more common forms of CHD constituting approximately 7\% of all CHD’s \cite{Liu2019CHD}.
CoA is defined as the narrowing of a point within the aorta, usually at the isthmus, which may be discrete or elongated and occurs on a spectrum with varying severity. This narrowing leads  to  increased  resistance  across  the  vessel  thus causing  upper  body  hypertension  and  reduced blood  supply  to  downstream  vessels  and  body  parts.  The  resultant  pressure  difference  across  the coarctation is used by clinicians as a means of diagnosis, and is referred to as the pressure gradient. A  peak  systolic  pressure  difference  of  greater  than  20 mmHg or diagnosis via imaging following hypertension, warrants intervention \cite{zuhlke2019congenital,obaid2019computer}.  Common  treatments  for  CoA  are  balloon  angioplasty, resection  with  end-to-end anastomosis (REEA) and stenting in older patients \cite{zuhlke2019congenital}. The overall objective of treatment is to alleviate the pressure difference and restore normal flow across the coarctation. However, recoarctation may occur  post  intervention  and the  haemodynamic  environment  post-surgery is a strong driver of this occurrence \cite{cecchi2011recoarctation}. Computational  fluid  dynamics  (CFD)  has  been  identified  as  a  potential  avenue  for  studying  patient haemodynamics and identifying key flow metrics such as pressure, velocity, and wall shear stress, with high temporal (time) and spatial resolution \cite{swanson2020patient}. Another key  benefit is the non-invasive nature of the technique. Furthermore, with the design of patient specific CFD pipelines, flow modelling can be specific to an individual patient. However, incorporating CFD within clinical workflows has been limited due to challenges in patient data acquisition for modelling and the skill set required to conduct the modelling \cite{capelli2018patient}. Thus, most applications of the technique have mainly resided in research.

In resource constrained settings the design of patient-specific CFD pipelines that are clinically applicable requires careful consideration. Patient-specific pipelines by their nature require data from each individual patient. The nature and complexity of the model therefore depends on data availability and acquisition, via a specific imaging modality. 

In their 2019 study, Swanson et al \cite{swanson2020patient} conducted a patient specific CoA CFD study that used computed tomography angiography (CTA) and doppler transthoracic echocardiography (Doppler TTE)  as   the   geometry  and   velocity  data   acquisition   modalities,   respectively,   and an open source  CFD  solver \cite{swanson2020patient}.  The  goal  of  the  study  was  to  develop  a  pipeline  that  could  be implemented in resource constrained settings \cite{swanson2020patient}. In addition, they demonstrated the suitability of doppler echocardiography as a velocity data acquisition tool and motivated the case for further study of its implementation. This, despite current preference for phase contrast magnetic resonance imaging (PC-MRI) in patient-specific CFD literature \cite{pc_mri_1,pc_mri_2, pc_mri_3, pc_mri_4, pc_mri_5}. Although PC-MRI is an adequate modality from a technical standpoint, the devices tend to be costly. Furthermore, its implementation within the paediatric population is challenging, as it may require that the patients are sedated, intubated and ventilated for the procedure, to enable breath holding, which carries risks. 

Swanson et al highlighted the current challenges using doppler  echocardiography in a CFD study. One of the main concerns was the variations in patient heart rate across different measurements, which is an inherent limitation of the modality. This discrepancy arises because measurements across each vessel must be taken at different points in time. As a result, the velocity profile at the inlet and outlets did not have the same heart rate. To remedy this, the velocity profiles across each of the vessels were scaled to the same heart rate of 120BPM. This assumption, which considered the impact of heart rate on velocity profiles as negligible introduced errors.

Machine learning, is a family of methods that are used to automatically discover patterns within data \cite{bishop2016pattern}. They have proven to be effective function approximators in cases where an analytical expression is difficult to obtain but data is available. In CFD, ML has been typically used to decrease the high computational cost associated with traditional numerical flow solution calculations \cite{feiger2020accelerating,nita2022personalized,yevtushenko2021deep}. This has varied between solely training an ML model to compute flow parameters to having a hybrid approach where data-driven approaches are combined with numerical methods. The general approach to coupling ML with CFD has been to generate data using numerical simulations and training the ML model on this data. 

Yevtushenko et al \cite{yevtushenko2021deep} demonstrated how Deep Learning could be used to develop a centre-aggregated aortic haemodynamics model, to compute parameters such as pressure gradient in CoA. They consider their approach as an alternative to numerical modelling in cases where low computational power and time are of high importance. While their approach uses ML as the sole method for computing flow, Nita et al \cite{nita2022personalized} take a hybrid approach in which they embed a ML model to compute pressure gradients in CoA, within their proposed reduced order multi-scale model. Another subclass of approaches are physics based ML methods, where physical laws such as mass continuity and the Navier-Stokes equations are embedded into the structure of the model. Raissi et al \cite{raissi2019pinn}, for example, introduce physics informed neural networks and demonstrate their ability to solve forward and inverse problems involving nonlinear partial differential equations. These studies illustrate the variety of ways in which ML methods can be applied to physics based problems relating to flow modelling in CoA.

Various studies have shown the need for accurate BCs given their effect on CFD modelling results \cite{madhavan2018effect,morbiducci2013inflow,pirola2017choice}. Our paper develops a framework that uses a ML approach to improve Doppler TTE-derived boundary conditions (BCs) for use in a CFD model of CoA. By accounting for changes in the velocity profiles, due to changes in heart rate, more accurate inlet and outlet BCs can be computed. An ML approach using non-Artificial Neural Network (non-ANN) methods was selected as our dataset was relatively small and we we were looking to find patterns or relationships within the data \cite{hands-on}. Given the widespread use of echocardiography in a wide range of clinical settings, our approach strengthens its potential for use in patient-specific CFD pipelines by deriving realistic BCs across all vessels. The paper begins by describing the development of the ML-CFD pipeline. The ML subsection describes model training and the subsequent CFD section focuses on the modelling approach and validation process. The results, discussion and conclusion are then presented.

\section{Methods: ML-CFD Pipeline Development}\label{sec2}
\begin{figure}[H]
\centering
\includegraphics[width=0.8\textwidth]{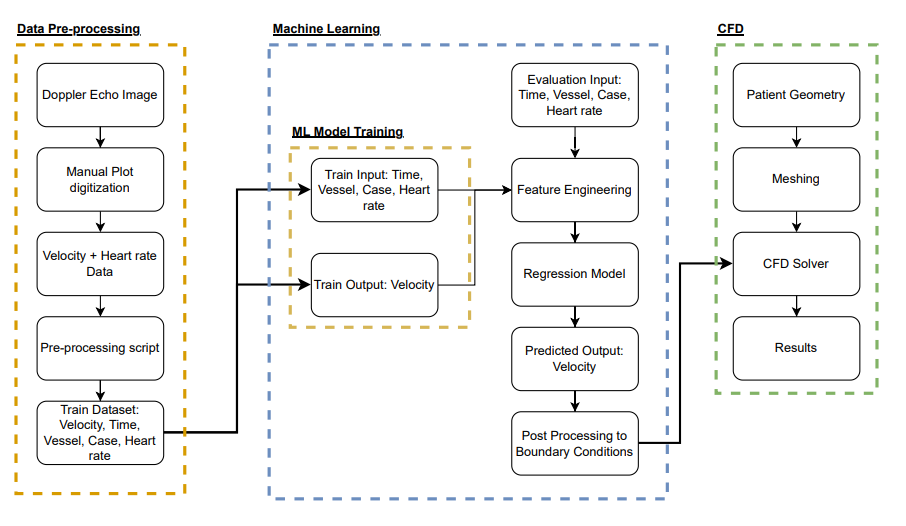}
\caption{\label{fig:Method Pipeline} Flow Diagram representation of the full framework used to develop the coupled ML-CFD Pipeline, for modelling patient-specific flow using Doppler TTE as the velocity data acquisition modality}
\end{figure}

In this section, we outline the methods used to develop our approach. Figure \ref{fig:Method Pipeline} provides an overview of the full pipeline. Beginning with data pre-procssing, key features are extracted from the Doppler TTE images to build the ML training dataset. This dataset is split into the respective inputs and outputs, as shown in the ML Model Training box, which then undergo a feature engineering process. This converts the data into a format that is compatible with the ML model. Once the regression model is trained, the evaluation inputs are specified and BCs obtained. These are then input to the CFD solver along with meshed patient geometry. The CFD solver then outputs the velocity and pressure results.  

\subsection{Data Pipeline and Clinical Data}\label{subsec2}
In this study Doppler TTE images for both pre- and post- intervention across the ascending aorta, supra aortic branches (innominate artery, left common carotid artery and left subclavian artery), coarctation and descending aorta, and a 3D patient-specific geometry were used, as obtained from \cite{swanson2020patient}. In addition, four images across the coarctation constituting 1 pre- and 3 post- intervention cases were provided by Red Cross War Memorial Children's Hospital (RXH). 

A data pre-processing pipeline was developed to process the Doppler TTE images and build a dataset that could be used to train the ML models.

The velocities were extracted from the Doppler TTE images using a free online plot digitizer \cite{plotdigitizer}. This was a manual process conducted by the researcher. Points were first set along the axes of the velocity graph to convert the image to a plot.The values at the designated points were then entered manually, which then scaled the plot accordingly. Points were then traced along the velocity profile within the graph to obtain the velocity readings. The tracings were done for velocity profiles that were adequately developed, that is, the distinct shape of the profile was visible and there was minimal noise across the profile. Tracings during diastole were set to zero. The coordinates of the tracings were then saved in a csv file. This process was repeated for all the Doppler TTE images in the dataset from  \cite{swanson2020patient} and for the 4 additional coarctation images from RXH.

The same plot digitizer was used to determine the number of beats per second, enabling determination of the heart rate of each profile. The clinicians noted that measuring the heart rate directly from the image was a more reliable approach. Thus, in images where the electrocardiogram (ECG) was present, the heart rate was calculated by measuring the time interval between successive peaks of the QRS complex. However, in images where the ECG was not present, the time interval between the successive peaks of the velocity profiles was used to calculate the heart rate. In both cases, the time intervals between 4-5 successive peaks were measured and the average calculated, which was then used as the associated heart rate. We note that this approach introduces human error, especially in cases where the ECG was not present. Given that the sampling rate at which the velocity profiles were measured was not available, methods such as using a Fourier transform were not an option. As such it would not have been possible to find the right scaling for the y-axis. 

The csv files from the plot digitizer were then processed using an in-house python script to build the dataset for ML model training. Firstly, the script zeroed the velocity profiles such that they started from 0 and inverted the coarctation velocity values to the positive y-axis. Secondly, any non 0 values during diastole were set to 0 and the velocity profile was interpolated for 200 or 350 steps within the time interval, depending on the length of the time interval and number of cycles. Lastly, the associated vessel name (ascending aorta, innominate artery, left common carotid artery, left subclavian artery, coarctation and descending aorta), case (pre or post intervention), and heart rate were captured . A training dataset was then generated from this information in the form of a table.

\subsection{Machine Learning Pipeline}\label{subsec2}

Our training dataset consisted of 3 numerical values, namely time in seconds, velocity in metres per second and heart rate in beats per second.The Case and Vessel features/ columns comprised of categorical or non-numeric entries. Additional pre-processing steps were employed before ML model training. Firstly, the categorical entries Case and Vessel were converted to numeric entries using Scikit-learns label encoding function \cite{sklearn}, as the models within the Scikit-learns library require numerical inputs. For the numeric entries (in particular the velocity values) it was found that the distribution of the data was skewed to the right and contained several outliers. Data points which were set to 0 during diastole were partially to blame for the skewing of the distribution, resulting in several outliers in the box and whisker plot. High velocity values at the tail end were due to the coarctation. This would contribute to poor performance for models that are sensitive to outliers and lead to higher error on ML model predictions. However, these outliers were biologically plausible and not as a result of noise or measurement errors, Numpy's log1p function a Log transform was applied to the velocity entries to reduce the effect of the outliers during training  \cite{numpy}, and improve ML model performance while preserving the outlier values within training.

\begin{figure}[H]
     \centering
     \begin{subfigure}[b]{0.48\textwidth}
         \centering
         \includegraphics[width=\textwidth]{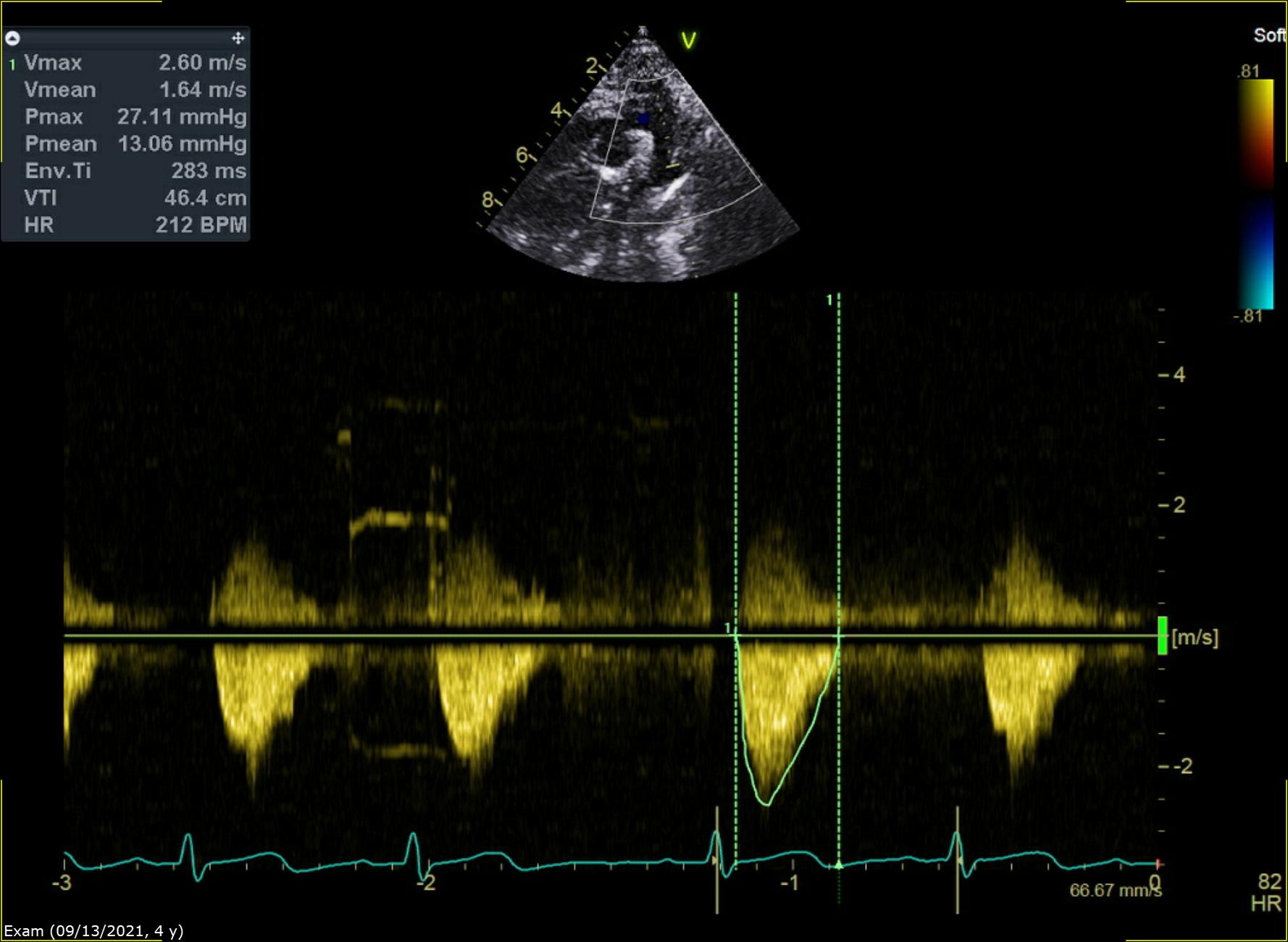}
         \caption{Coarctation measurement post intervention}
         \label{fig:AIM3ML post}
     \end{subfigure}
     \hfill
     \begin{subfigure}[b]{0.45\textwidth}
         \centering
         \includegraphics[width=\textwidth]{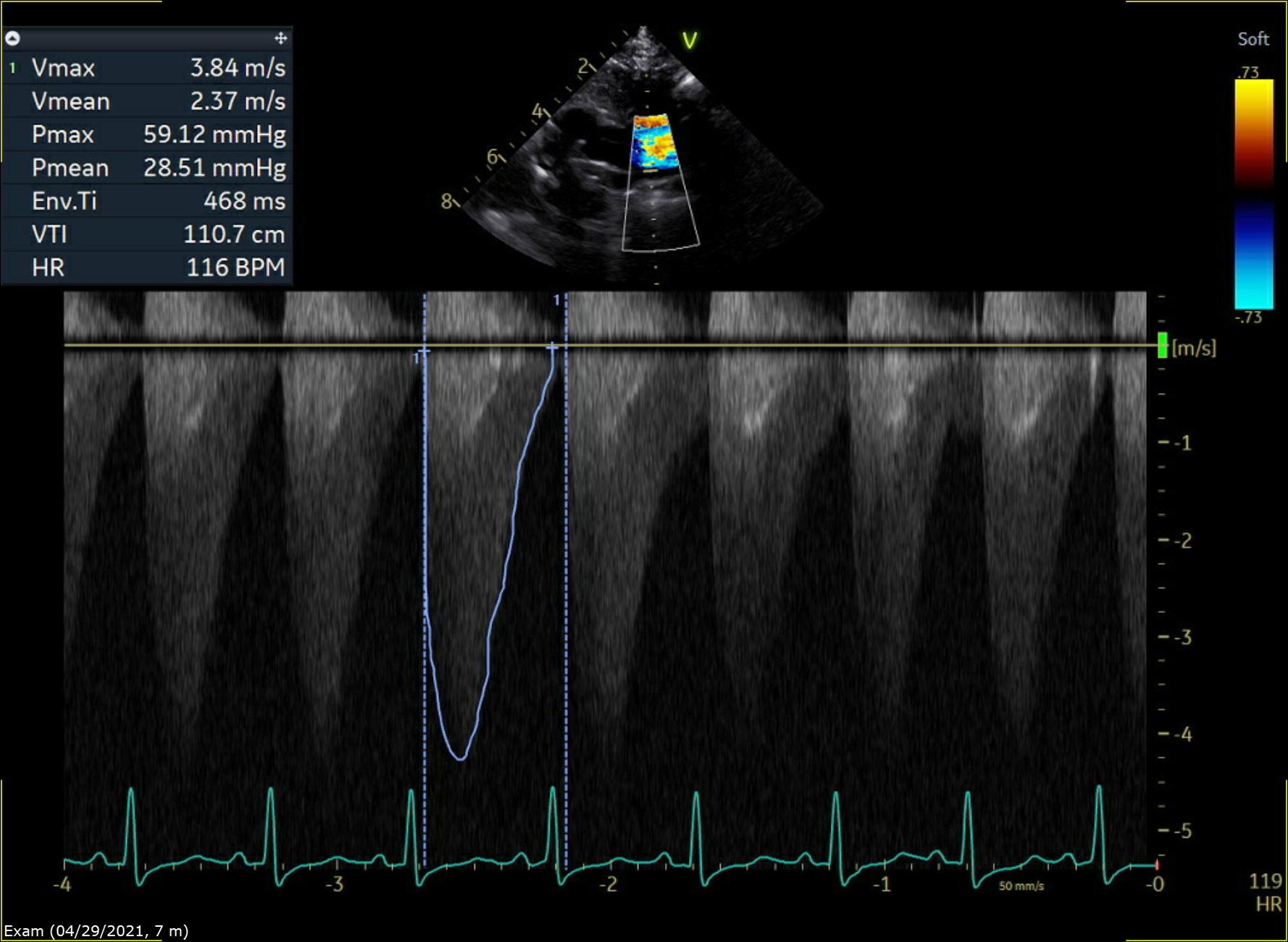}
         \caption{Coarctation measurement pre intervention}
         \label{fig:AIM3ML pre}
     \end{subfigure}
     \newline
     \hfill
     \begin{subfigure}[b]{0.45\textwidth}
         \centering
         \includegraphics[width=\textwidth]{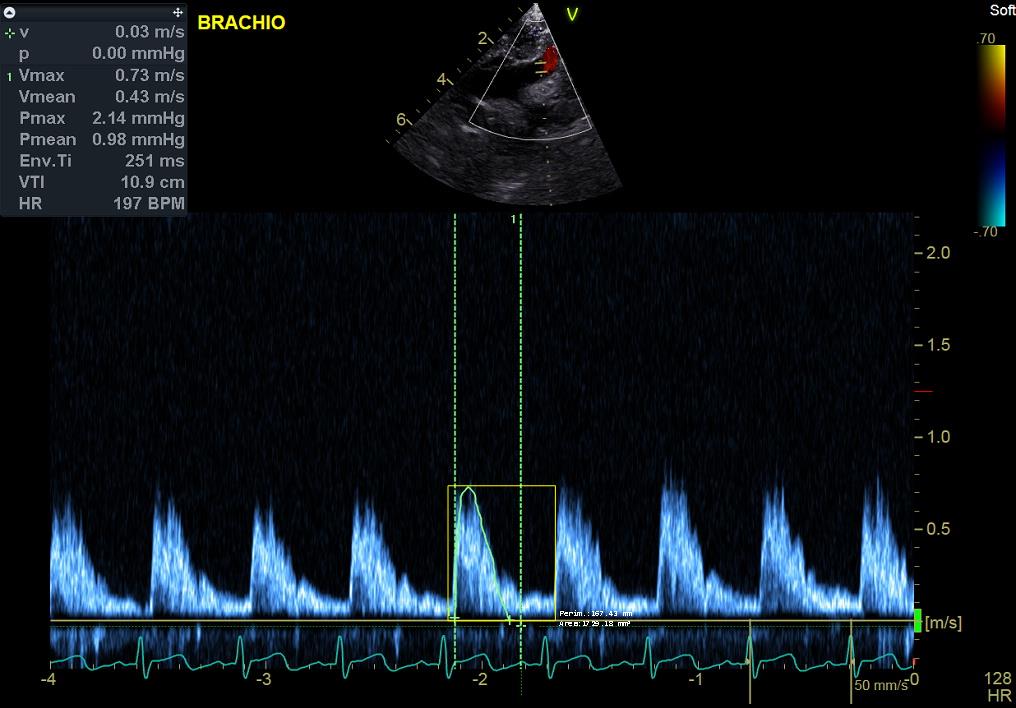}
         \caption{Innominate Artery measurement post intervention}
         \label{fig:Liam Innom post}
     \end{subfigure}
      \hfill
     \begin{subfigure}[b]{0.45\textwidth}
         \centering
         \includegraphics[width=\textwidth]{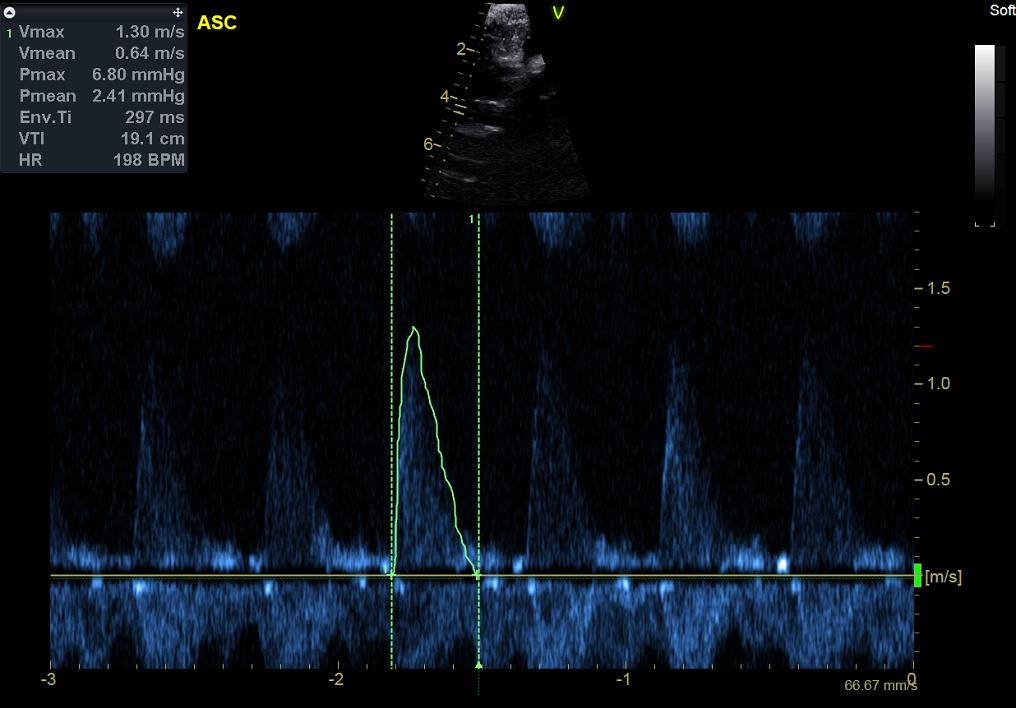}
         \caption{Ascending Aorta measurement pre intervention}
         \label{fig:Liam Asc pre}
     \end{subfigure}
        \caption{Sample Doppler TTE images that were used within the study}
        \label{fig:Doppler TTE images}
\end{figure}

The class of ML problem posed is a regression problem as the objective of the ML model is to predict a continuous value given a set of discrete inputs. Recalling that the main challenge faced, the discontinuity in the time of measurement between vessels, the heart rate at the current measured vessel varied from the previous vessel. It was noted earlier that changes in the heart rate affect the velocity profile across the measured vessel, which indicates a present relationship between heart rate and velocity. For the CFD model, a mass flow rate is prescribed at the inlets and outlets of the region of interest and is a function of velocity. Therefore, the patients heart rate was seen as the key input that could relate the time-dependent in-vivo flow conditions with the velocity, which was then used to calculate the BCs. Thus, time (time interval over which the velocity was measured), velocity (velocity across the measured vessel site) and the heart rate (average heart rate across 3-4 cardiac cycles) were the identified key numerical inputs. However, these values changed depending on the vessel and the case (whether an intervention had occurred or not). These two parameters were therefore included as categorical variables. It is noted that although the vessel name which represents anatomical position can be better represented by geometrical parameters which could convey more insightful information, this in our case was not possible. For geometric information was inconsistent within the given data, whilst the vessel name was consistently given. 

For this study the inputs to the ML models were time in seconds, the vessel name (ascending aorta, innominate artery, left common carotid artery, left subclavian artery, coarctation and descending aorta), case (pre or post intervention) and heart rate in beats per second. Each of these inputs were scaled using Scikit-learns Standard Scaler function \cite{sklearn}. This was to limit error in models sensitive to scaling, such as linear regression and support vector machines. The evaluation metric used to assess model performance was the root mean square error (RMSE) \cite{hands-on}. The models trained in this study were, namely; Linear Regression model, which provided the baseline, Support Vector Machine Regressor, Random Forest Regressor, Gradient Boosted Regressor, K-Neighbours Regressor (k-NN) and a voting ensemble method. The train dataset was split into a train, validation and test with a 60\%, 20\%, and 20\% split, respectively. Parameters for the K-neighbours regressor, random forest regressor, gradient boosted regressor, LightGBM regressor \cite{ke2017lightgbm} and support vector machine regressor were optimized on the train set and validated against the validation set \cite{sklearn}. Finally, all models were fit to the combined training and validation set, and evaluated on the test set.

\subsection{CFD Pipeline}\label{subsec2}
ANSYS Fluent (ANSYS Academic Research [Fluent], release 2020R2) was the commercial software package used for the CFD study. A mesh constituting 2 million elements was generated usng the unrepaired geometry in \cite{swanson2020patient}. The number of elements used were greater than the minimum number of elements required for grid independence. The incompressible Navier-Stokes equations were solved using the finite volume method. While the pressure was computed using the PRESTO discretization scheme, the momentum was determined using a 2\textsuperscript{nd} order upwind scheme. Pressure and velocity were coupled using the PISO scheme. For temporal discretization, a bounded 2\textsuperscript{nd} order implicit scheme was used with a time step of 0.05ms. Flow was modeled as laminar and blood was assumed to be Newtonian, with a density of 1060 kg/m\textsuperscript{3} and viscosity of  0.004 kg/(m s). The vessel wall was modelled as rigid and a no-slip boundary condition was applied. Convergence criteria of 10\textsuperscript{-3} was used (ANSYS Academic Research [Fluent], release 2020R2, 37.21.1, Judging Convergence, ANSYS, Inc.).

The maximum coarctation velocity obtained from the numerical simulation was compared to the measured value from Doppler TTE. This was achieved, by calculating the percentage difference between the two values. As noted in \cite{swanson2020patient}, Doppler TTE is capable of reliably measuring the velocity of blood within the body and the measurement was taken by an experienced sonographer. This measurement was therefore used to validate the numerical solution. Although, pressure is the clinically relevant flow parameter, the simplified Bernoulli method used by the device has been shown to be inaccurate. Therefore, velocity was a more reliable flow parameter for validating the framework.  

To calculate the BCs, 5 ML models with the least RMSE were selected to generate BCs for the CFD modelling process. A time interval of 0.442 seconds, which corresponds to 135.6BPM, was used. This was the calculated heart rate at which the measurement across the coarctation was taken. The vessel names and case (pre- intervention) were also input to each ML model. The CFD mass flow rate input profiles were calculated from the velocity profiles output by each ML model using, the stated density of blood and the vessel area. The set of BCs were adapted from \cite{pirola2017choice,madhavan2018effect,swanson2020patient} with the aim of assessing BCs that were commonly used in literature, whilst also taking into account the limited data constraint due to the use of Doppler TTE. Each set of BCs had to be simulated for the not-adjusted case (max velocity values obtained from Doppler TTE measurements with no additional processing) and for the outputs of the selected 5 ML models. The computational cost of running the simulations was therefore an important factor.

Table \ref{BCs} shows the set of BCs specified at the patient geometry inlet and outlets for each ML model and the not-adjusted case. The outlets consisted of the arch branches (AoB) (innominate artery (outlet 1); left common carotid artery, (LCCA, outlet 2); left subclavian artery, (LSA, outlet 3)) and descending aorta (DAo, outlet 4). The inlet was defined at the ascending aorta. Four different boundary condition combinations were explored. For these combinations, the inlet BC was kept constant while the outlet BCs varied, as these BCs had greater impact on flow within the coarctation \cite{madhavan2018effect}. BC 1, which specified zero pressure at all outlets was shown to result in the least accurate flow solution when compared to other types of outlet BCs \cite{pirola2017choice}. Thus, BC 1 was primarily included as a benchmark case to provide the baseline performance of the ML-CFD pipeline. The zero pressure BC with target mass flow adds a specified mass flow rate as a constraint that must be met when calculating pressure at the outlet. This provided an alternate way for pressure to be prescribed at the outlet, as mass flow rate can be seen as an alternate form of prescribing velocity at the outlets. As can be seen in BC 3 and BC 4.

Given that Madhavan et al showed that the effects of the inlet boundary condition are only significant up to 2 inlet diameters distal to the inlet patch \cite{madhavan2018effect}, a plug flow velocity profile was assumed for the inlet. The inlet is sufficiently far from the coarctation, which is the main region of interest. The effect of outlet BCs is registered up to 5 diameters distal to the outlet \cite{madhavan2018effect}, hence greater care was taken with their definition. Due to limited data availability imposed by using Doppler TTE, methods such as the 3 Element Windkessel model (3-EWM) could not be applied. Instead, simpler outlet BCs namely zero pressure and mass flow rate, were used. The ML model derived BCs were compared with the not-adjusted BCs, which were the BCs obtained from Doppler TTE in \cite{swanson2020patient}. Differences in the BCs used in the study were largely due to the magnitudes of the values used.

\begin{table}[h]
\begin{flushleft}
\begin{tabular}{llll}    
\toprule
BC & Inlet BC & AoB BC & DAo\\
\midrule
BC 1 & mass flow rate & zero pressure & zero pressure\\
BC 2 & mass flow rate & mass flow rate & zero pressure\\
BC 3 & mass flow rate & zero pressure, target mass flow rate & zero pressure, target mass flow rate\\
BC 4 & mass flow rate & mass flow rate & zero pressure, target mass flow rate\\
\botrule
\end{tabular}
\caption{Boundary Conditions: This table contains the 4 sets of boundary conditions (BCs) that were used within the study, namely BC 1-4. Outlets are located at the arch branches (AoB) and the descending aorta, with the inlet at the ascending aorta. The types of BCs used comprised of mass flow rate [kg/s], zero pressure (static pressure [Pa] set to zero) and zero pressure target mass flow rate (static pressure [Pa] set to 0 initially but constrained to meet a specified mass flow [kg/s] value in subsequent iterations).
}
\label{BCs}
\end{flushleft}
\end{table}

\section{Results}\label{sec3}
This section begins with a brief description of the data processing results, followed by the ML model results and finally the CFD results.

\subsection{Data Pipeline and Clinical Data Results}\label{subsec2}
The data processing pipeline output is a training dataset of size (3650 x 5). The 5 columns are titled; time in seconds, velocity in meters per second, case (pre or post intervention), vessel and the heart rate in beats per second. 

\subsection{ML Model Training Results}\label{subsec2}
Table \ref{ML Model Results} shows the RMSE values for the ML models that were trained and evaluated on the test set. Amongst the models, the k-NN regressor is found to have the best fit as it had the lowest RMSE. The voting regressor which is an ensemble of the k-NN and random forest regressors, performed second best. An interesting observation is that the top 5 models primarily comprise of tree-based methods, with k-NN being the exception. However, it must be noted that the RMSE is not the final metric with which the ML models are evaluated. Instead the comparison of maximum velocity across the coarctation, obtained from ANSYS Fluent, compared to the ground truth value obtained by Doppler TTE is also used. The support vector machine regressor and linear regression had the poorest results, as they had the highest RMSE values.

\begin{table}[h]
\begin{center}
\begin{minipage}{174pt}
\begin{tabular}{ll}                
\toprule
Model & RMSE[m/s] \\
\midrule
Linear Regression    & 0.4169     \\
Support Vector Machine Regressor & 0.3695   \\
Gradient Boosted Regressor & 0.2678 \\
Light GBM Regressor & 0.1194 \\
Random Forest Regressor & 0.0604\\
Random Forest Regressor Optimized & 0.0585\\
Voting Regressor & 0.0500 \\
K-Nearest Neighbours Regressor    &  0.0493   \\
\botrule
\end{tabular}
\caption{ML Model Results: This table presents the root mean square error (RSME) values for each machine learning model, when evaluated against the test or hold out set. The test set constituted 20\% of the original dataset built for the study.}\label{ML Model Results}%
\end{minipage}
\end{center}
\end{table}

\subsection{Boundary Condition Results}\label{subsec2}

\begin{figure}[H]
\centering
\includegraphics[width=0.8\textwidth]{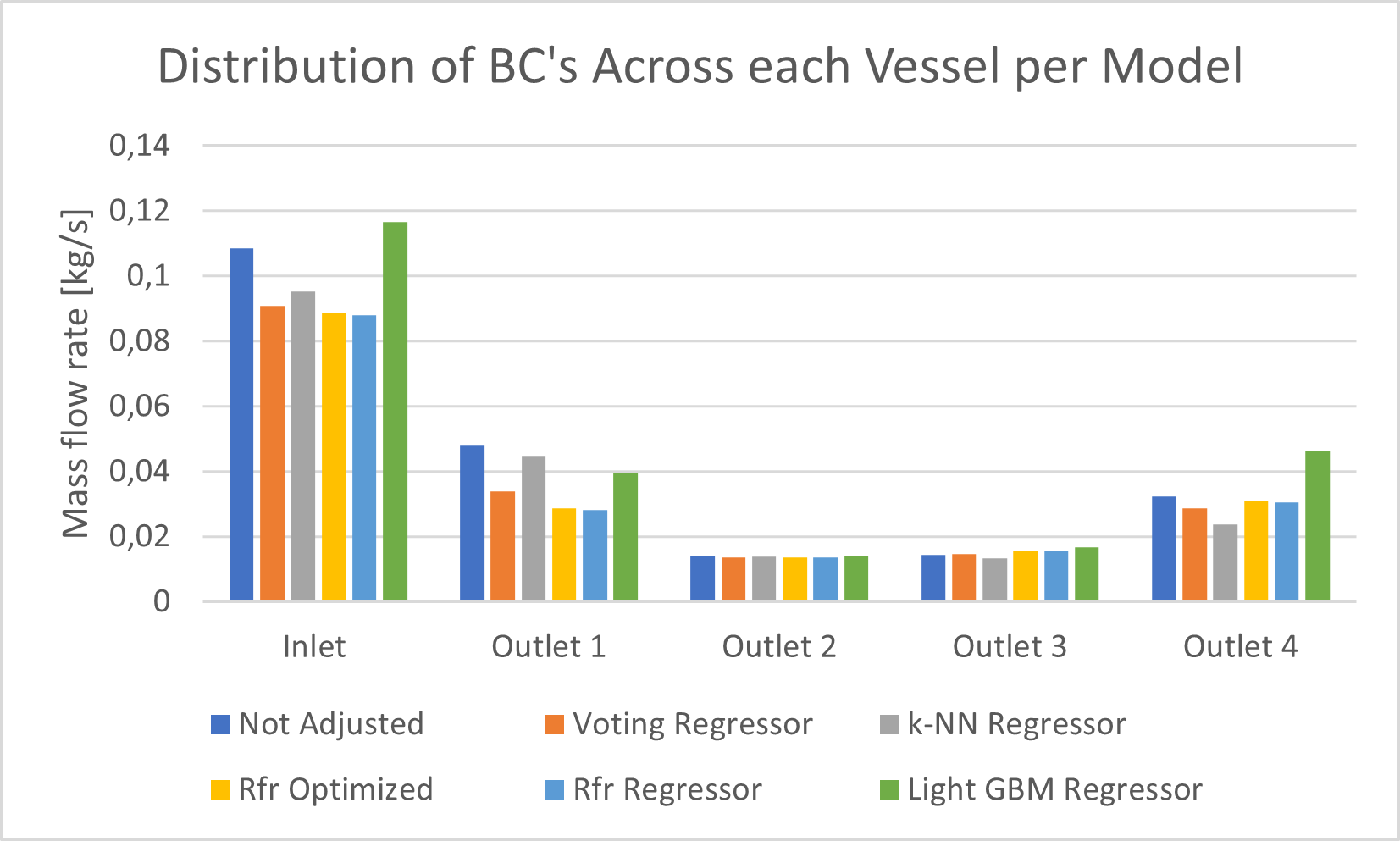}
\caption{\label{fig:ML model BCs across each vessel}Outlets 1 to 4 refer to the innominate artery, left common carotid artery, left subclavian artery and descending aorta, respectively. We firstly observe that the ML models output BCs whose magnitudes are physiologically reasonable. At outlets 2 and 3 there is close agreement between the ML models and the not-adjusted case, with percent deviations of 1\% and 6\%, respectively. At outlets 1 and 4 less agreement is seen as higher percent deviations are observed at 19\% and 15\%, respectively. Similarly, higher variation is observed at the inlet but it is less than that at outlets 1 and 4. With regards to the ML models, there are three notable trends. First, the LightGBM Regressor overestimates the mass flow across the vessel except at outlet 1. Secondly, the voting ensemble and Random Forest regressors (hyperparameter tuned and default) output values with similar magnitudes. Whilst the k-NN regressor across some vessels underestimates or over estimates values output by the voting ensemble and Random Forest regressors.}
\end{figure}

From Figure \ref{fig:ML model BCs across each vessel} we observe that we are able to obtain reasonable mass flow rate values from each of the ML models. There is close agreement between the ML model BCs at outlets 2 and 3, which on average deviate from the mean by  1\% and 6\%, respectively. Similarly, outlet 4 showed  good agreement among the model BCs with the exception of the LightGBM Regressor, which overestimated the mass flow across the outlet. At outlet 1, all the ML model BCs underestimate the flow in comparison to the not-adjusted BC. It is at this outlet that we observe the highest variation amongst the models at 19\%. This is followed by the inlet where the second highest variation is observed at 10\%. The LightGBM Regressor outputs the highest mass flow rate, with the remaining ML models outputting values less than that of the not-adjusted case. 

\subsection{Coarctation Velocity Results}\label{subsec2}

\begin{figure}[H]
     \centering
     \begin{subfigure}[b]{\textwidth}
         \centering
         \includegraphics[width=\textwidth]{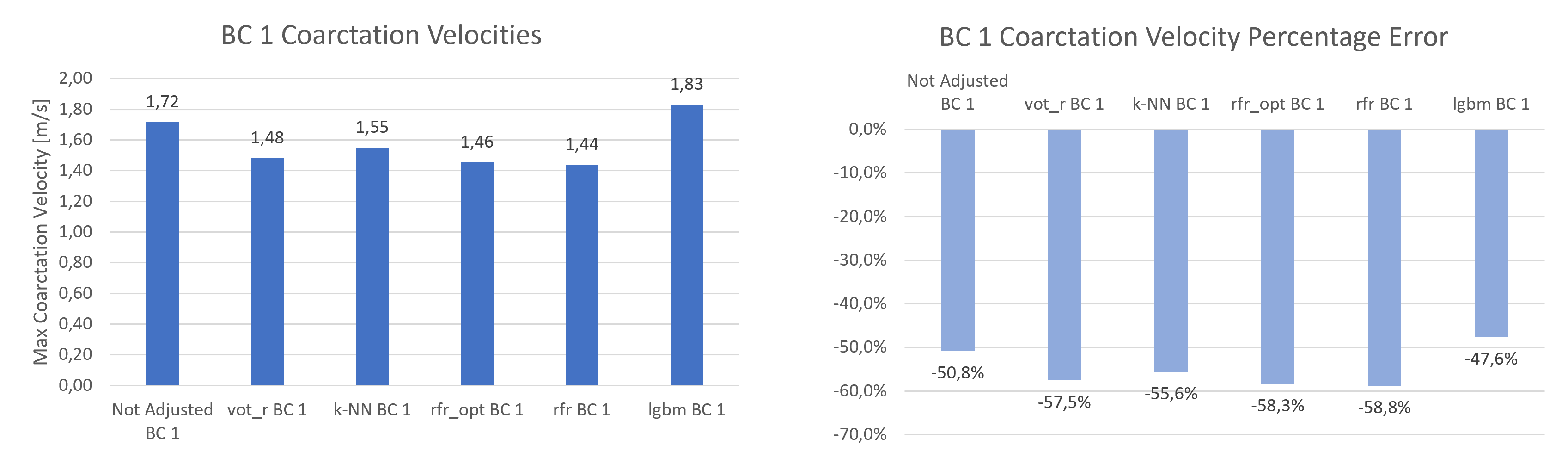}
         \caption{BC 1 results}
         \label{fig:BC 1 results}
     \end{subfigure}
     \newline
     \hfill
     \begin{subfigure}[b]{\textwidth}
         \centering
         \includegraphics[width=\textwidth]{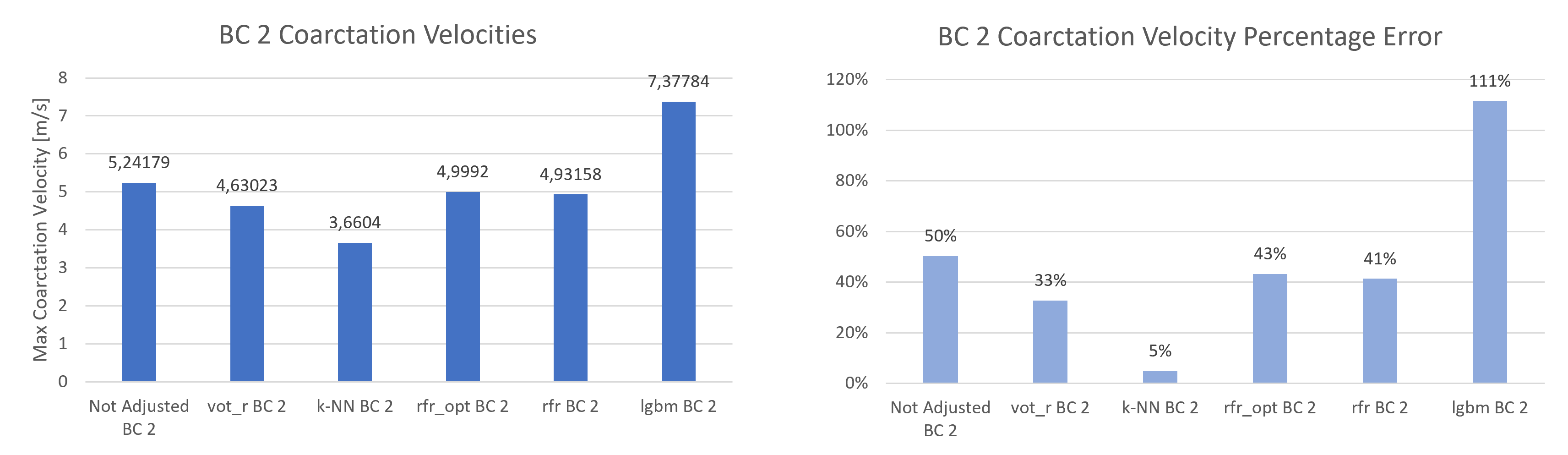}
         \caption{BC 2 results}
         \label{fig:BC 1 results}
     \end{subfigure}
     \newline
     \hfill
     \begin{subfigure}[b]{\textwidth}
         \centering
         \includegraphics[width=\textwidth]{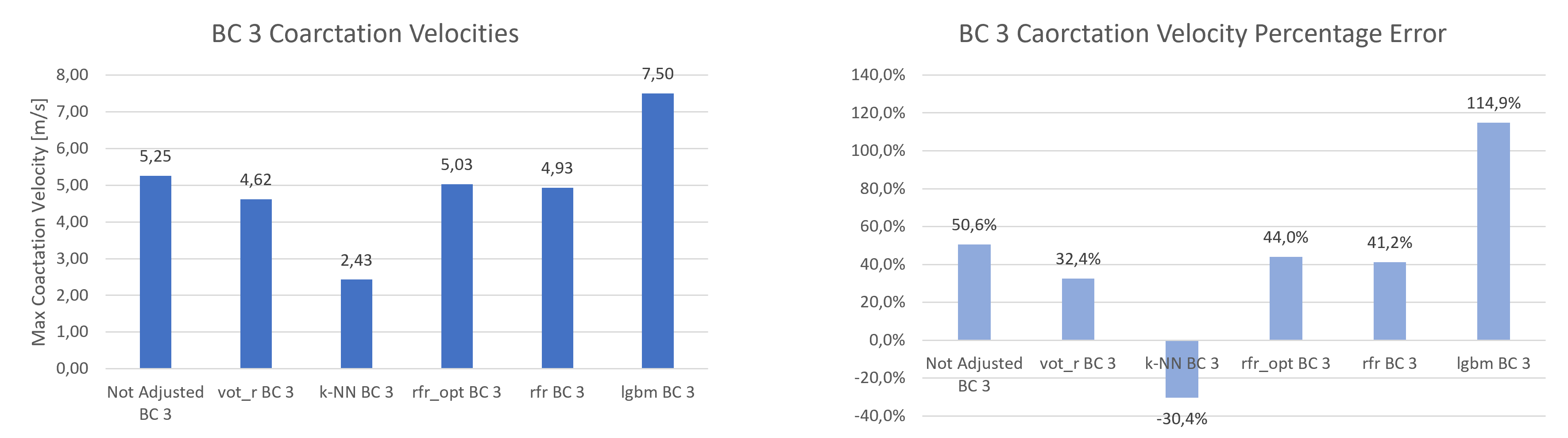}
         \caption{BC 3 results}
         \label{fig:BC 3 results}
     \end{subfigure}
     \newline
      \hfill
     \begin{subfigure}[b]{\textwidth}
         \centering
         \includegraphics[width=\textwidth]{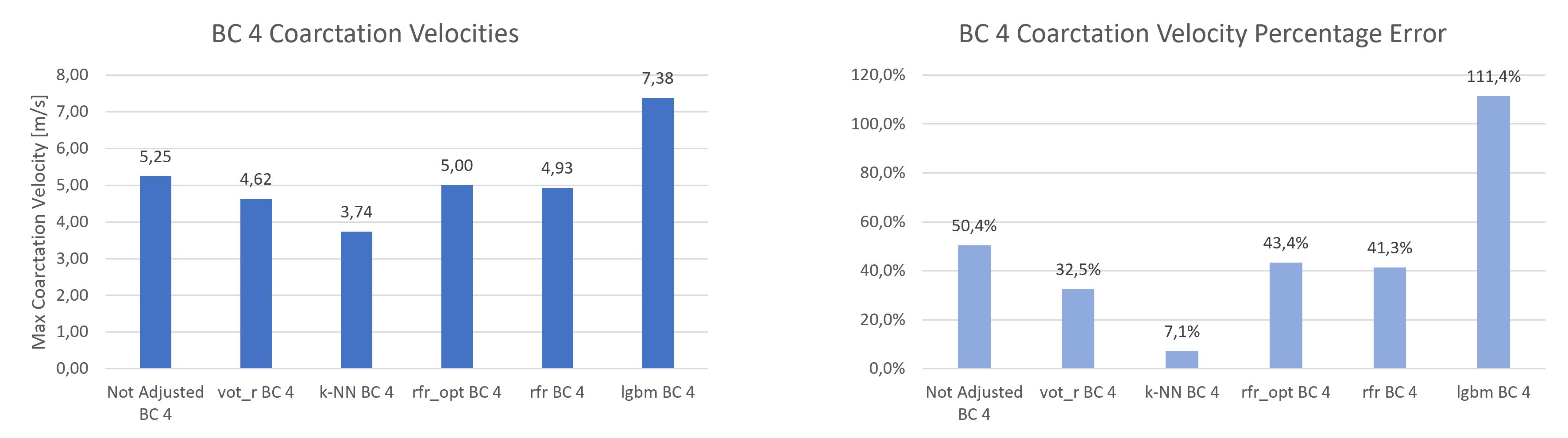}
         \caption{BC 4 results}
         \label{fig:BC 4 results}
     \end{subfigure}
        \caption{Results for each type of boundary condition (BC) used.}
        \label{fig:Coarctation Results}
\end{figure}


Figure \ref{fig:Coarctation Results} presents the maximum coarctation velocities and percentage error (the percentage difference between the ground truth max coarctation velocity 3.49 [m/s] and the output of the simulation) for each BC type and model used. At first glance, it can be seen that the performance of the ML models varied among BC 1-4, in Table \ref{BCs}. Of the ML derived BCs, we observe that the k-NN model BCs were found to have the least percentage error, with BC 2 having the lowest value of 4.88\%. Similarly the k-NN regressor using BC 4 output the second most accurate result at 7\%. The remaining ML models generally performed poorly across all BCs, with the voting ensemble regressor showing the least percentage error outside that of the k-NN regressor. BC 1 (zero pressure outlet BCs) resulted in the least accurate results except for the case of the LightGBM regressor. For this BC, only the LightGBM showed marginal improvement as compared to the not-adjusted case. Between BC 2 and 3 the voting ensemble, Random Forest (hyperparameter tuned and default) and LightGBM regressor's performance did not vary significantly. However, the k-NN regressor's performance varied significantly in BC 1 and 3.

\section{Discussion and Conclusion}\label{sec4}
Numerical modelling offers a means for improved analysis and understanding of cardiovascular haemodynamics, through simulation of flow in complex geometries. However, the inputs required for these methods can prove challenging to obtain, particularly in resource constrained settings. This study aimed to add to the body of work at the intersection of empirically derived patient-specific BCs, doppler echocardiography, CFD and ML \cite{pant2014methodological,lassila2020population,swanson2020patient}.  

The framework is capable of producing velocity results that are physiologically accurate, whilst accounting for variations in the patients' heart rate. This is evidenced by percentage differences within 5\% for BC2 and 7\% for BC4 using the k-NN regressor. It is noteworthy that the k-NN performed the best among the ML models. Apart from Linear Regression, it is the simplest amongst the ML models. In addition, this may suggest there is a relationship between the input variables and the outputs based on euclidean distance. The poor performance of the tree-based methods (voting ensemble, Random Forest, and LightGBM regressor's) suggests that the underlying relationship that maps the input variables to the output may not be achieved by asking a series of binary questions, which is the underlying assumption used by tree-based models. The poor performance of the tree based methods also highlights that a low RMSE value may not result in an accurate flow solution. Therefore, a more suitable metric for ML model selection prior to calculating flow using CFD will be required. However, it justifies the validation approach taken within the study, namely that the CFD model velocity field solution was the final measure by which the framework was validated. 

The results suggest that mass flow rate outlet BCs show potential as a feasible preliminary alternative to pressure based 3-EWM which are traditionally favored within literature \cite{pant2014methodological,morbiducci2013inflow,obaid2019computer}, in cases where pressure data is absent and doppler echocardiography is used. Even though 3-EWM has proven capable of prescribing physiologically accurate BCs that result in highly accurate flow solutions, it requires accurate patient pressure data and calibration of model parameters such as resistance and capacitance. Although there are approaches such as the one presented in \cite{pant2014methodological}, which are capable of calibrating 3-EWM parameters in cases where limited data are available, pressure data are still required. Thus, such an approach would have proved difficult to implement in this study. It would be possible to obtain parameters from literature with similar studies, however, this would still introduce errors due to variability between patients. Therefore, this approach opens the door for haemodynamic modelling not only in cases where Doppler TTE is used as the velocity data acquisition modality but also those where pressure data may not be available. Noting the limitation of pressure data, future work would be to extend the framework to include 3-EWM given the strong motivation within literature for the approach.

\subsection{Limitations}\label{subsec2}
The dataset built for this study was limited as it was restricted to one patient and the 4 additional Doppler TTE images that were provided. Therefore, the generalizability of the ML models is limited. Extrapolating for cases beyond those used within the study, may result in a performance drop of the ML-CFD pipeline. Furthermore, the study was limited in the type of cases modelled. Modelling for intra-patient variability, for example when the patient is in a stressed state, was not possible. Therefore, modelling was limited to a range of heart rates that capture a relaxed state. Both \cite{lassila2020population} and \cite{pant2014methodological} note that there is a significant difference between flow parameters for the 2 different states, with \cite{pant2014methodological} further noting increased difficulty in accurately modelling flow. Similarly, the effects of different patient geometries on the performance of the framework were not evaluated. Among the inputs used to train the ML models, the relative importance between vessel names was not taken into account when label encoding was used. Lastly, the maximum coarctation velocities results show that the choice of ML model used (and the type of BC prescribed at the outlet) had an impact on the outcomes. This motivates the need to evaluate the performance of other models such as artificial neural networks (ANNs) that may be able to determine more complex relationships within the data \cite{feiger2020accelerating,yevtushenko2021deep,raissi2019pinn}.

Assumptions made in the CFD modelling also contributed to the limitations of the study. These include, the use of fixed BCs and the rigid wall assumption. However, accounting for these considerations comes at increased computational cost, hence their inclusion should be balanced against improvements in accuracy \cite{pant2014methodological}. A suitable compromise could be including these factors in the ML model component. In their paper, \cite{rackauckas2020universal} motivate for the application and advantage of universal partial differential equations, which is a hybrid approach that combines physics based and ML models. Thus, future work could include the development of a hybrid ML and physics based lumped parameter model that accounts for the noted limitations.

\bmhead{Supplementary information}
\bmhead{Acknowledgments}
Computations were performed using facilities provided by the University of Cape Town’s ICTS High Performance Computing team: hpc.uct.ac.za

\section*{Declarations}

\begin{itemize}
\item Conflict of interest: The authors declare that they have no conflict of interest. 
\item Ethics approval: 
The study used secondary data from the Red Cross War Memorial Children's Hospital. Patient consent was obtained.
\item Code availability: The code used within this study is available at the given link: https://figshare.com/s/a86bf491ef0a5b1e38cb
\item Authors' contributions: VMMP contributed to the design of the pipeline, methodology, analysis and writing of the manuscript. MN, AKM contributed to the conceptualization of the work, analysis, methodology and critical revision of the manuscript. TA, JL and LZ contributed to data collection and critical revision of the manuscript. 
\end{itemize}

\bigskip

\begin{appendices}

\section{Boundary Condition Calculations}\label{secA1}

This appendix contains additional details on how the boundary conditions were calculated to enable easy reproducibility of the framework discussed within this paper. 

The boundary conditions were calculated by first identifying the time at which peak flow at the inlet occurred. This time value was then used to obtain the mass flow rate values that occurred at this time. The initial value at the inlet was then replaced by the sum of the outlet mass flow rates in order to conserve mass continuity. This was firstly because within the body due to the compliance of the vessels mass is not conserved and the models were trained on physiological data the conservation law would not be learned. Secondly, doing this resulted in values which were less erroneous as conservation of mass is an assumption made by the ANSYS Fluent pressure based solver (ANSYS Academic Research [Fluent], release 2020R2). Figures below show the BC values used to generate the CFD results in this study. The abbreviations mf, zp and zpt are mass flow rate [kg/s], zero pressure and zero pressure with targeted mass flow rate, respectively.

\begin{figure}[H]
\centering
\includegraphics[width=0.8\textwidth]{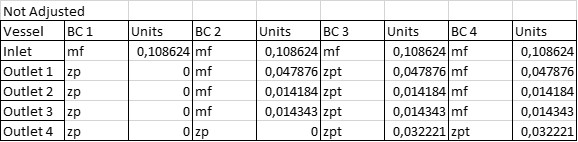}
\caption{\label{fig:NotAdj_BCs}Not-Adjusted BC Values}
\end{figure}

\begin{figure}[H]
\centering
\includegraphics[width=0.8\textwidth]{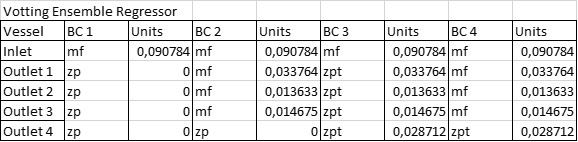}
\caption{\label{fig:vot_r_BCs}Voting Ensemble BC Values}
\end{figure}

\begin{figure}[H]
\centering
\includegraphics[width=0.8\textwidth]{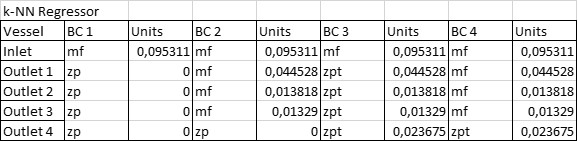}
\caption{\label{fig:kNN_BCs}k-NN BC Values }
\end{figure}

\begin{figure}[H]
\centering
\includegraphics[width=0.8\textwidth]{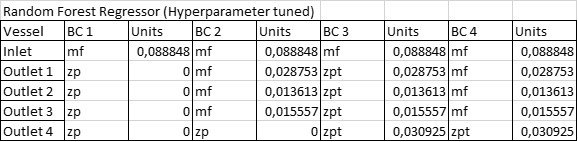}
\caption{\label{fig:rfr_opt_BCs}Random Forest (Hyperparameter tuned) BC Values}
\end{figure}

\begin{figure}[H]
\centering
\includegraphics[width=0.8\textwidth]{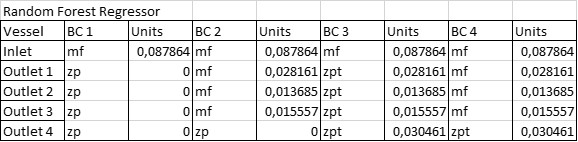}
\caption{\label{fig:rfr_BCs}Random Forest BC Values}
\end{figure}

\begin{figure}[H]
\centering
\includegraphics[width=0.8\textwidth]{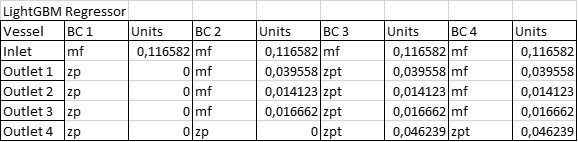}
\caption{\label{fig:lgbm_BCs}LightGBM BC Values}
\end{figure}




\end{appendices}



\begin{thebibliography}{27}
\ifx \bisbn   \undefined \def \bisbn  #1{ISBN #1}\fi
\ifx \binits  \undefined \def \binits#1{#1}\fi
\ifx \bauthor  \undefined \def \bauthor#1{#1}\fi
\ifx \batitle  \undefined \def \batitle#1{#1}\fi
\ifx \bjtitle  \undefined \def \bjtitle#1{#1}\fi
\ifx \bvolume  \undefined \def \bvolume#1{\textbf{#1}}\fi
\ifx \byear  \undefined \def \byear#1{#1}\fi
\ifx \bissue  \undefined \def \bissue#1{#1}\fi
\ifx \bfpage  \undefined \def \bfpage#1{#1}\fi
\ifx \blpage  \undefined \def \blpage #1{#1}\fi
\ifx \burl  \undefined \def \burl#1{\textsf{#1}}\fi
\ifx \doiurl  \undefined \def \doiurl#1{\url{https://doi.org/#1}}\fi
\ifx \betal  \undefined \def \betal{\textit{et al.}}\fi
\ifx \binstitute  \undefined \def \binstitute#1{#1}\fi
\ifx \binstitutionaled  \undefined \def \binstitutionaled#1{#1}\fi
\ifx \bctitle  \undefined \def \bctitle#1{#1}\fi
\ifx \beditor  \undefined \def \beditor#1{#1}\fi
\ifx \bpublisher  \undefined \def \bpublisher#1{#1}\fi
\ifx \bbtitle  \undefined \def \bbtitle#1{#1}\fi
\ifx \bedition  \undefined \def \bedition#1{#1}\fi
\ifx \bseriesno  \undefined \def \bseriesno#1{#1}\fi
\ifx \blocation  \undefined \def \blocation#1{#1}\fi
\ifx \bsertitle  \undefined \def \bsertitle#1{#1}\fi
\ifx \bsnm \undefined \def \bsnm#1{#1}\fi
\ifx \bsuffix \undefined \def \bsuffix#1{#1}\fi
\ifx \bparticle \undefined \def \bparticle#1{#1}\fi
\ifx \barticle \undefined \def \barticle#1{#1}\fi
\bibcommenthead
\ifx \bconfdate \undefined \def \bconfdate #1{#1}\fi
\ifx \botherref \undefined \def \botherref #1{#1}\fi
\ifx \url \undefined \def \url#1{\textsf{#1}}\fi
\ifx \bchapter \undefined \def \bchapter#1{#1}\fi
\ifx \bbook \undefined \def \bbook#1{#1}\fi
\ifx \bcomment \undefined \def \bcomment#1{#1}\fi
\ifx \oauthor \undefined \def \oauthor#1{#1}\fi
\ifx \citeauthoryear \undefined \def \citeauthoryear#1{#1}\fi
\ifx \endbibitem  \undefined \def \endbibitem {}\fi
\ifx \bconflocation  \undefined \def \bconflocation#1{#1}\fi
\ifx \arxivurl  \undefined \def \arxivurl#1{\textsf{#1}}\fi
\csname PreBibitemsHook\endcsname

\bibitem{Liu2019CHD}
\begin{barticle}
\bauthor{\bsnm{Liu}, \binits{Y.}},
\bauthor{\bsnm{Chen}, \binits{S.}},
\bauthor{\bsnm{Zühlke}, \binits{L.}},
\bauthor{\bsnm{Black}, \binits{G.C.}},
\bauthor{\bsnm{Choy}, \binits{M.-k.}},
\bauthor{\bsnm{Li}, \binits{N.}},
\bauthor{\bsnm{Keavney}, \binits{B.D.}}:
\batitle{{Global birth prevalence of congenital heart defects 1970–2017:
  updated systematic review and meta-analysis of 260 studies}}.
\bjtitle{International Journal of Epidemiology}
\bvolume{48}(\bissue{2}),
\bfpage{455}--\blpage{463}
(\byear{2019})
{\href{https://arxiv.org/abs/https://academic.oup.com/ije/article-pdf/48/2/455/28479664/dyz009.pdf}{{https://academic.oup.com/ije/article-pdf/48/2/455/28479664/dyz009.pdf}}}.
\doiurl{10.1093/ije/dyz009}
\end{barticle}
\endbibitem

\bibitem{zuhlke2019congenital}
\begin{barticle}
\bauthor{\bsnm{Z{\"u}hlke}, \binits{L.}},
\bauthor{\bsnm{Lawrenson}, \binits{J.}},
\bauthor{\bsnm{Comitis}, \binits{G.}},
\bauthor{\bsnm{De~Decker}, \binits{R.}},
\bauthor{\bsnm{Brooks}, \binits{A.}},
\bauthor{\bsnm{Fourie}, \binits{B.}},
\bauthor{\bsnm{Swanson}, \binits{L.}},
\bauthor{\bsnm{Hugo-Hamman}, \binits{C.}}:
\batitle{Congenital heart disease in low-and lower-middle--income countries:
  current status and new opportunities}.
\bjtitle{Current cardiology reports}
\bvolume{21}(\bissue{12}),
\bfpage{1}--\blpage{13}
(\byear{2019})
\end{barticle}
\endbibitem

\bibitem{obaid2019computer}
\begin{barticle}
\bauthor{\bsnm{Obaid}, \binits{D.R.}},
\bauthor{\bsnm{Smith}, \binits{D.}},
\bauthor{\bsnm{Gilbert}, \binits{M.}},
\bauthor{\bsnm{Ashraf}, \binits{S.}},
\bauthor{\bsnm{Chase}, \binits{A.}}:
\batitle{Computer simulated “virtual tavr” to guide tavr in the presence of
  a previous starr-edwards mitral prosthesis}.
\bjtitle{Journal of Cardiovascular Computed Tomography}
\bvolume{13}(\bissue{1}),
\bfpage{38}--\blpage{40}
(\byear{2019})
\end{barticle}
\endbibitem

\bibitem{cecchi2011recoarctation}
\begin{barticle}
\bauthor{\bsnm{Cecchi}, \binits{E.}},
\bauthor{\bsnm{Giglioli}, \binits{C.}},
\bauthor{\bsnm{Valente}, \binits{S.}},
\bauthor{\bsnm{Lazzeri}, \binits{C.}},
\bauthor{\bsnm{Gensini}, \binits{G.F.}},
\bauthor{\bsnm{Abbate}, \binits{R.}},
\bauthor{\bsnm{Mannini}, \binits{L.}}:
\batitle{Role of hemodynamic shear stress in cardiovascular disease}.
\bjtitle{Atherosclerosis}
\bvolume{214}(\bissue{2}),
\bfpage{249}--\blpage{256}
(\byear{2011}).
\doiurl{10.1016/j.atherosclerosis.2010.09.008}
\end{barticle}
\endbibitem

\bibitem{swanson2020patient}
\begin{barticle}
\bauthor{\bsnm{Swanson}, \binits{L.}},
\bauthor{\bsnm{Owen}, \binits{B.}},
\bauthor{\bsnm{Keshmiri}, \binits{A.}},
\bauthor{\bsnm{Deyranlou}, \binits{A.}},
\bauthor{\bsnm{Aldersley}, \binits{T.}},
\bauthor{\bsnm{Lawrenson}, \binits{J.}},
\bauthor{\bsnm{Human}, \binits{P.}},
\bauthor{\bsnm{De~Decker}, \binits{R.}},
\bauthor{\bsnm{Fourie}, \binits{B.}},
\bauthor{\bsnm{Comitis}, \binits{G.}}, \betal:
\batitle{A patient-specific cfd pipeline using doppler echocardiography for
  application in coarctation of the aorta in a limited resource clinical
  context}.
\bjtitle{Frontiers in Bioengineering and Biotechnology}
\bvolume{8},
\bfpage{409}
(\byear{2020})
\end{barticle}
\endbibitem

\bibitem{capelli2018patient}
\begin{barticle}
\bauthor{\bsnm{Capelli}, \binits{C.}},
\bauthor{\bsnm{Sauvage}, \binits{E.}},
\bauthor{\bsnm{Giusti}, \binits{G.}},
\bauthor{\bsnm{Bosi}, \binits{G.M.}},
\bauthor{\bsnm{Ntsinjana}, \binits{H.}},
\bauthor{\bsnm{Carminati}, \binits{M.}},
\bauthor{\bsnm{Derrick}, \binits{G.}},
\bauthor{\bsnm{Marek}, \binits{J.}},
\bauthor{\bsnm{Khambadkone}, \binits{S.}},
\bauthor{\bsnm{Taylor}, \binits{A.M.}}, \betal:
\batitle{Patient-specific simulations for planning treatment in congenital
  heart disease}.
\bjtitle{Interface Focus}
\bvolume{8}(\bissue{1}),
\bfpage{20170021}
(\byear{2018})
\end{barticle}
\endbibitem

\bibitem{pc_mri_1}
\begin{barticle}
\bauthor{\bsnm{Pirola}, \binits{S.}},
\bauthor{\bsnm{Jarral}, \binits{O.}},
\bauthor{\bsnm{O'Regan}, \binits{D.}},
\bauthor{\bsnm{Asimakopoulos}, \binits{G.}},
\bauthor{\bsnm{Anderson}, \binits{J.}},
\bauthor{\bsnm{Pepper}, \binits{J.}},
\bauthor{\bsnm{Athanasiou}, \binits{T.}},
\bauthor{\bsnm{Xu}, \binits{X.}}:
\batitle{Computational study of aortic hemodynamics for patients with an
  abnormal aortic valve: The importance of secondary flow at the ascending
  aorta inlet}.
\bjtitle{APL bioengineering}
\bvolume{2}(\bissue{2}),
\bfpage{026101}
(\byear{2018})
\end{barticle}
\endbibitem

\bibitem{pc_mri_2}
\begin{barticle}
\bauthor{\bsnm{Antonuccio}, \binits{M.N.}},
\bauthor{\bsnm{Mariotti}, \binits{A.}},
\bauthor{\bsnm{Fanni}, \binits{B.M.}},
\bauthor{\bsnm{Capellini}, \binits{K.}},
\bauthor{\bsnm{Capelli}, \binits{C.}},
\bauthor{\bsnm{Sauvage}, \binits{E.}},
\bauthor{\bsnm{Celi}, \binits{S.}}:
\batitle{Effects of uncertainty of outlet boundary conditions in a
  patient-specific case of aortic coarctation}.
\bjtitle{Annals of biomedical engineering}
\bvolume{49}(\bissue{12}),
\bfpage{3494}--\blpage{3507}
(\byear{2021})
\end{barticle}
\endbibitem

\bibitem{pc_mri_3}
\begin{barticle}
\bauthor{\bsnm{Br{\"u}ning}, \binits{J.}},
\bauthor{\bsnm{Hellmeier}, \binits{F.}},
\bauthor{\bsnm{Yevtushenko}, \binits{P.}},
\bauthor{\bsnm{K{\"u}hne}, \binits{T.}},
\bauthor{\bsnm{Goubergrits}, \binits{L.}}:
\batitle{Uncertainty quantification for non-invasive assessment of pressure
  drop across a coarctation of the aorta using cfd}.
\bjtitle{Cardiovascular engineering and technology}
\bvolume{9}(\bissue{4}),
\bfpage{582}--\blpage{596}
(\byear{2018})
\end{barticle}
\endbibitem

\bibitem{pc_mri_4}
\begin{botherref}
\oauthor{\bsnm{Campbell}, \binits{I.C.}},
\oauthor{\bsnm{Ries}, \binits{J.}},
\oauthor{\bsnm{Dhawan}, \binits{S.S.}},
\oauthor{\bsnm{Quyyumi}, \binits{A.A.}},
\oauthor{\bsnm{Taylor}, \binits{W.R.}},
\oauthor{\bsnm{Oshinski}, \binits{J.N.}}:
Effect of inlet velocity profiles on patient-specific computational fluid
  dynamics simulations of the carotid bifurcation.
Journal of biomechanical engineering
\textbf{134}(5)
(2012)
\end{botherref}
\endbibitem

\bibitem{pc_mri_5}
\begin{barticle}
\bauthor{\bsnm{Conti}, \binits{M.}},
\bauthor{\bsnm{Long}, \binits{C.}},
\bauthor{\bsnm{Marconi}, \binits{M.}},
\bauthor{\bsnm{Berchiolli}, \binits{R.}},
\bauthor{\bsnm{Bazilevs}, \binits{Y.}},
\bauthor{\bsnm{Reali}, \binits{A.}}:
\batitle{Carotid artery hemodynamics before and after stenting: A patient
  specific cfd study}.
\bjtitle{Computers \& Fluids}
\bvolume{141},
\bfpage{62}--\blpage{74}
(\byear{2016})
\end{barticle}
\endbibitem

\bibitem{bishop2016pattern}
\begin{bbook}
\bauthor{\bsnm{Bishop}, \binits{C.M.}}:
\bbtitle{Pattern Recognition and Machine Learning}.
\bsertitle{Information Science and Statistics}.
\bpublisher{Springer}, \blocation{???}
(\byear{2016}).
\burl{https://books.google.co.za/books?id=kOXDtAEACAAJ}
\end{bbook}
\endbibitem

\bibitem{feiger2020accelerating}
\begin{barticle}
\bauthor{\bsnm{Feiger}, \binits{B.}},
\bauthor{\bsnm{Gounley}, \binits{J.}},
\bauthor{\bsnm{Adler}, \binits{D.}},
\bauthor{\bsnm{Leopold}, \binits{J.A.}},
\bauthor{\bsnm{Draeger}, \binits{E.W.}},
\bauthor{\bsnm{Chaudhury}, \binits{R.}},
\bauthor{\bsnm{Ryan}, \binits{J.}},
\bauthor{\bsnm{Pathangey}, \binits{G.}},
\bauthor{\bsnm{Winarta}, \binits{K.}},
\bauthor{\bsnm{Frakes}, \binits{D.}}, \betal:
\batitle{Accelerating massively parallel hemodynamic models of coarctation of
  the aorta using neural networks}.
\bjtitle{Scientific reports}
\bvolume{10}(\bissue{1}),
\bfpage{1}--\blpage{13}
(\byear{2020})
\end{barticle}
\endbibitem

\bibitem{nita2022personalized}
\begin{barticle}
\bauthor{\bsnm{Nita}, \binits{C.-I.}},
\bauthor{\bsnm{Puiu}, \binits{A.}},
\bauthor{\bsnm{Bunescu}, \binits{D.}},
\bauthor{\bsnm{Mihai~Itu}, \binits{L.}},
\bauthor{\bsnm{Mihalef}, \binits{V.}},
\bauthor{\bsnm{Chintalapani}, \binits{G.}},
\bauthor{\bsnm{Armstrong}, \binits{A.}},
\bauthor{\bsnm{Zampi}, \binits{J.}},
\bauthor{\bsnm{Benson}, \binits{L.}},
\bauthor{\bsnm{Sharma}, \binits{P.}}, \betal:
\batitle{Personalized pre-and post-operative hemodynamic assessment of aortic
  coarctation from 3d rotational angiography}.
\bjtitle{Cardiovascular Engineering and Technology}
\bvolume{13}(\bissue{1}),
\bfpage{14}--\blpage{40}
(\byear{2022})
\end{barticle}
\endbibitem

\bibitem{yevtushenko2021deep}
\begin{botherref}
\oauthor{\bsnm{Yevtushenko}, \binits{P.}},
\oauthor{\bsnm{Goubergrits}, \binits{L.}},
\oauthor{\bsnm{Gundelwein}, \binits{L.}},
\oauthor{\bsnm{Setio}, \binits{A.}},
\oauthor{\bsnm{Heimann}, \binits{T.}},
\oauthor{\bsnm{Ramm}, \binits{H.}},
\oauthor{\bsnm{Lamecker}, \binits{H.}},
\oauthor{\bsnm{Kuehne}, \binits{T.}},
\oauthor{\bsnm{Meyer}, \binits{A.}},
\oauthor{\bsnm{Schafstedde}, \binits{M.}}:
Deep learning based centerline-aggregated aortic hemodynamics: An efficient
  alternative to numerical modelling of hemodynamics.
IEEE Journal of Biomedical and Health Informatics
(2021)
\end{botherref}
\endbibitem

\bibitem{raissi2019pinn}
\begin{barticle}
\bauthor{\bsnm{Raissi}, \binits{M.}},
\bauthor{\bsnm{Perdikaris}, \binits{P.}},
\bauthor{\bsnm{Karniadakis}, \binits{G.E.}}:
\batitle{Physics-informed neural networks: A deep learning framework for
  solving forward and inverse problems involving nonlinear partial differential
  equations}.
\bjtitle{Journal of Computational Physics}
\bvolume{378},
\bfpage{686}--\blpage{707}
(\byear{2019}).
\doiurl{10.1016/j.jcp.2018.10.045}
\end{barticle}
\endbibitem

\bibitem{madhavan2018effect}
\begin{barticle}
\bauthor{\bsnm{Madhavan}, \binits{S.}},
\bauthor{\bsnm{Kemmerling}, \binits{E.M.C.}}:
\batitle{The effect of inlet and outlet boundary conditions in image-based cfd
  modeling of aortic flow}.
\bjtitle{Biomedical engineering online}
\bvolume{17}(\bissue{1}),
\bfpage{1}--\blpage{20}
(\byear{2018})
\end{barticle}
\endbibitem

\bibitem{morbiducci2013inflow}
\begin{barticle}
\bauthor{\bsnm{Morbiducci}, \binits{U.}},
\bauthor{\bsnm{Ponzini}, \binits{R.}},
\bauthor{\bsnm{Gallo}, \binits{D.}},
\bauthor{\bsnm{Bignardi}, \binits{C.}},
\bauthor{\bsnm{Rizzo}, \binits{G.}}:
\batitle{Inflow boundary conditions for image-based computational hemodynamics:
  impact of idealized versus measured velocity profiles in the human aorta}.
\bjtitle{Journal of biomechanics}
\bvolume{46}(\bissue{1}),
\bfpage{102}--\blpage{109}
(\byear{2013})
\end{barticle}
\endbibitem

\bibitem{pirola2017choice}
\begin{barticle}
\bauthor{\bsnm{Pirola}, \binits{S.}},
\bauthor{\bsnm{Cheng}, \binits{Z.}},
\bauthor{\bsnm{Jarral}, \binits{O.}},
\bauthor{\bsnm{O'Regan}, \binits{D.}},
\bauthor{\bsnm{Pepper}, \binits{J.}},
\bauthor{\bsnm{Athanasiou}, \binits{T.}},
\bauthor{\bsnm{Xu}, \binits{X.}}:
\batitle{On the choice of outlet boundary conditions for patient-specific
  analysis of aortic flow using computational fluid dynamics}.
\bjtitle{Journal of biomechanics}
\bvolume{60},
\bfpage{15}--\blpage{21}
(\byear{2017})
\end{barticle}
\endbibitem

\bibitem{hands-on}
\begin{bbook}
\bauthor{\bsnm{G\`{e}ron}, \binits{A.}}:
\bbtitle{Hands-on Machine Learning with Scikit-Learn and TensorFlow : Concepts,
  Tools, and Techniques to Build Intelligent Systems}.
\bpublisher{O'Reilly Media},
\blocation{Sebastopol, CA}
(\byear{2017})
\end{bbook}
\endbibitem

\bibitem{plotdigitizer}
\begin{botherref}
PlotDigitizer: Version 2.2
(2022).
\url{https://plotdigitizer.com}
\end{botherref}
\endbibitem

\bibitem{sklearn}
\begin{barticle}
\bauthor{\bsnm{Pedregosa}, \binits{F.}},
\bauthor{\bsnm{Varoquaux}, \binits{G.}},
\bauthor{\bsnm{Gramfort}, \binits{A.}},
\bauthor{\bsnm{Michel}, \binits{V.}},
\bauthor{\bsnm{Thirion}, \binits{B.}},
\bauthor{\bsnm{Grisel}, \binits{O.}},
\bauthor{\bsnm{Blondel}, \binits{M.}},
\bauthor{\bsnm{Prettenhofer}, \binits{P.}},
\bauthor{\bsnm{Weiss}, \binits{R.}},
\bauthor{\bsnm{Dubourg}, \binits{V.}}, \betal:
\batitle{Scikit-learn: Machine learning in python}.
\bjtitle{Journal of machine learning research}
\bvolume{12}(\bissue{Oct}),
\bfpage{2825}--\blpage{2830}
(\byear{2011})
\end{barticle}
\endbibitem

\bibitem{numpy}
\begin{barticle}
\bauthor{\bsnm{Harris}, \binits{C.R.}},
\bauthor{\bsnm{Millman}, \binits{K.J.}},
\bauthor{\bparticle{van~der} \bsnm{Walt}, \binits{S.J.}},
\bauthor{\bsnm{Gommers}, \binits{R.}},
\bauthor{\bsnm{Virtanen}, \binits{P.}},
\bauthor{\bsnm{Cournapeau}, \binits{D.}},
\bauthor{\bsnm{Wieser}, \binits{E.}},
\bauthor{\bsnm{Taylor}, \binits{J.}},
\bauthor{\bsnm{Berg}, \binits{S.}},
\bauthor{\bsnm{Smith}, \binits{N.J.}},
\bauthor{\bsnm{Kern}, \binits{R.}},
\bauthor{\bsnm{Picus}, \binits{M.}},
\bauthor{\bsnm{Hoyer}, \binits{S.}},
\bauthor{\bparticle{van} \bsnm{Kerkwijk}, \binits{M.H.}},
\bauthor{\bsnm{Brett}, \binits{M.}},
\bauthor{\bsnm{Haldane}, \binits{A.}},
\bauthor{\bparticle{Fernández~del} \bsnm{Río}, \binits{J.}},
\bauthor{\bsnm{Wiebe}, \binits{M.}},
\bauthor{\bsnm{Peterson}, \binits{P.}},
\bauthor{\bsnm{Gérard-Marchant}, \binits{P.}},
\bauthor{\bsnm{Sheppard}, \binits{K.}},
\bauthor{\bsnm{Reddy}, \binits{T.}},
\bauthor{\bsnm{Weckesser}, \binits{W.}},
\bauthor{\bsnm{Abbasi}, \binits{H.}},
\bauthor{\bsnm{Gohlke}, \binits{C.}},
\bauthor{\bsnm{Oliphant}, \binits{T.E.}}:
\batitle{Array programming with {NumPy}}.
\bjtitle{Nature}
\bvolume{585},
\bfpage{357}--\blpage{362}
(\byear{2020}).
\doiurl{10.1038/s41586-020-2649-2}
\end{barticle}
\endbibitem

\bibitem{ke2017lightgbm}
\begin{botherref}
\oauthor{\bsnm{Ke}, \binits{G.}},
\oauthor{\bsnm{Meng}, \binits{Q.}},
\oauthor{\bsnm{Finley}, \binits{T.}},
\oauthor{\bsnm{Wang}, \binits{T.}},
\oauthor{\bsnm{Chen}, \binits{W.}},
\oauthor{\bsnm{Ma}, \binits{W.}},
\oauthor{\bsnm{Ye}, \binits{Q.}},
\oauthor{\bsnm{Liu}, \binits{T.-Y.}}:
Lightgbm: A highly efficient gradient boosting decision tree.
Advances in neural information processing systems
\textbf{30}
(2017)
\end{botherref}
\endbibitem

\bibitem{pant2014methodological}
\begin{barticle}
\bauthor{\bsnm{Pant}, \binits{S.}},
\bauthor{\bsnm{Fabr{\`e}ges}, \binits{B.}},
\bauthor{\bsnm{Gerbeau}, \binits{J.-F.}},
\bauthor{\bsnm{Vignon-Clementel}, \binits{I.}}:
\batitle{A methodological paradigm for patient-specific multi-scale cfd
  simulations: from clinical measurements to parameter estimates for individual
  analysis}.
\bjtitle{International journal for numerical methods in biomedical engineering}
\bvolume{30}(\bissue{12}),
\bfpage{1614}--\blpage{1648}
(\byear{2014})
\end{barticle}
\endbibitem

\bibitem{lassila2020population}
\begin{barticle}
\bauthor{\bsnm{Lassila}, \binits{T.}},
\bauthor{\bsnm{Sarrami-Foroushani}, \binits{A.}},
\bauthor{\bsnm{Hejazi}, \binits{S.}},
\bauthor{\bsnm{Frangi}, \binits{A.F.}}:
\batitle{Population-specific modelling of between/within-subject flow
  variability in the carotid arteries of the elderly}.
\bjtitle{International Journal for Numerical Methods in Biomedical Engineering}
\bvolume{36}(\bissue{1}),
\bfpage{3271}
(\byear{2020})
\end{barticle}
\endbibitem

\bibitem{rackauckas2020universal}
\begin{botherref}
\oauthor{\bsnm{Rackauckas}, \binits{C.}},
\oauthor{\bsnm{Ma}, \binits{Y.}},
\oauthor{\bsnm{Martensen}, \binits{J.}},
\oauthor{\bsnm{Warner}, \binits{C.}},
\oauthor{\bsnm{Zubov}, \binits{K.}},
\oauthor{\bsnm{Supekar}, \binits{R.}},
\oauthor{\bsnm{Skinner}, \binits{D.}},
\oauthor{\bsnm{Ramadhan}, \binits{A.}},
\oauthor{\bsnm{Edelman}, \binits{A.}}:
Universal differential equations for scientific machine learning.
arXiv preprint arXiv:2001.04385
(2020)
\end{botherref}
\endbibitem

\end{thebibliography}


\end{document}